\def\w0{\omega_0}

\documentclass{elsart}

\usepackage{epsfig}
\usepackage[dvips]{color}

\date{}

\begin{document}
\begin{frontmatter}

\title{Stochastic resonance and noise delayed extinction in a model of two competing species}
\author{D. Valenti\corauthref{cor}},
\corauth[cor]{Corresponding author}\ead{valentid@gip.dft.unipa.it}
\author{A. Fiasconaro} and \author{B. Spagnolo}
\address{Istituto Nazionale di Fisica della Materia, Unit\`a di Palermo and Dipartimento di Fisica e
Tecnologie Relative, University of Palermo Viale delle Scienze,
I-90128 Palermo, Italy}

\begin{abstract}
We study the role of the noise in the dynamics of two competing
species. We consider generalized Lotka-Volterra equations in the
presence of a multiplicative noise, which models the interaction
between the species and the environment. The interaction parameter
between the species is a random process which obeys a stochastic
differential equation with a generalized bistable potential in the
presence of a periodic driving term, which accounts for the
environment temperature variation. We find noise-induced periodic
oscillations of the species concentrations and stochastic
resonance phenomenon. We find also a nonmonotonic behavior of the
mean extinction time of one of the two competing species as a
function of the additive noise intensity.
\end{abstract}


\end{frontmatter}

\section{Introduction}\indent
Recently the influence of noise on population dynamics has been
the subject of intense theoretical investigations~
\cite{Ciu,Gia,Sch,Sta,Spa1}. The problem of the stability of the
biological complex systems in the presence of noise for example is
one of the more discussed in research activity of these last
years~\cite{Zim}. The noise through its interaction with the
nonlinearity of the systems can give rise to new counterintuitive
phenomena like stochastic resonance~\cite{Gam}, noise enhanced
stability~\cite{AguSpa}, noise delayed extinction, etc... The
principal aim of this work is the investigation of a generalized
Lotka-Volterra model with a random interaction parameter between
two competing species and in the presence of a multiplicative
noise~\cite{Ciu}. We find that the noise has a constructive role.
In fact it is responsible for the generation of quasi-periodic
temporal oscillations and the enhancement of the response of the
system to a driving force producing stochastic resonance. Moreover
we find noise delayed extinction, i. e. a nonmonotonic behavior of
the average extinction time of one of the two species as a
function of the noise intensity.
\section{The model}
Time evolution of two competing species is obtained within the
formalism of the Lotka-Volterra equations~\cite{Lot} in the
presence of a multiplicative noise

\begin{eqnarray}
\frac{dx}{dt}=\mu_1\thinspace x\thinspace(\alpha_1-x-\beta_1(t) y)+x\thinspace\xi_x(t)\\
\frac{dy}{dt}=\mu_2\thinspace y\thinspace(\alpha_2-y-\beta_2(t)
x)+y\thinspace\xi_y(t),
 \label{LotVol}
\end{eqnarray}
where $\xi_x(t)$ and $\xi_y(t)$ are statistically independent
Gaussian white noises with zero mean and correlation function
$\langle \xi_i(t)\xi_j(t')\rangle = \sigma
\delta(t-t')\delta_{ij}$ ($i,j=x,y$).
\begin{figure}[htbp]
\begin{center}
\includegraphics[width=9cm]{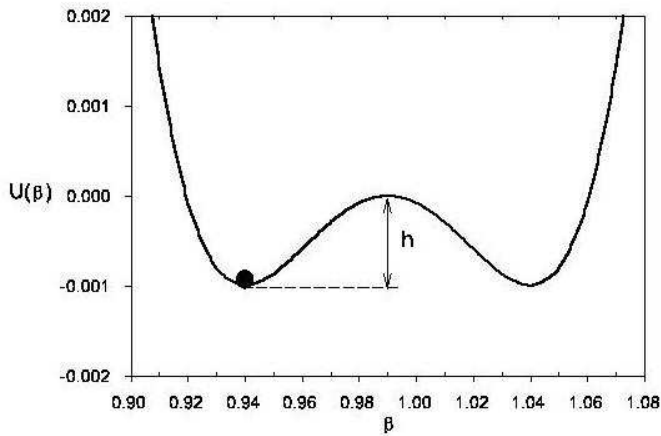}
\end{center}
\caption{ \small \emph{The bistable potential $U(\beta)$ of the
interaction parameter $\beta(t)$. The potential $U(\beta)$ is
centered on $\beta=0.99$. The parameters of the potential are
 $h = 6.25 \cdot 10^{-3}$, $\eta=0.05$, $\rho = -0.01$.}\bigskip}
 \label{potential}
\end{figure}
The time series for the two populations are obtained setting
$\alpha_1=\alpha_2=\alpha $, $\beta_1(t)=\beta_2(t)=\beta(t)$. It
is known that for $\beta < 1 $ a coexistence regime takes place,
that is both species survives, while for $\beta > 1 $ an exclusion
regime is established, that is one of the two species vanishes
after a certain time. Coexistence and exclusion of one of the two
species correspond to stable states of the Lotka-Volterra's
deterministic model~\cite{Baz}. Real ecosystems are immersed in a
noisy  nonstationary environment, so also the interaction
parameter is affected by the noise and some other deterministic
periodical driving such as the temperature. The change in the
competition rate between exclusion and coexistence e. g. occurs
randomly because of the coupling between the limiting resources
and the noisy environment. A random variation of limiting
resources produces a random competition between the species. The
noise therefore together with the periodic force determines the
crossing from a dynamical regime ($\beta < 1$, coexistence) to the
other one ($\beta > 1$, exclusion). To describe this continuous
and noisy behaviour of the interaction parameter $\beta(t)$ we
consider a stochastic differential equation with a bistable
potential and a periodical driving force

\begin{equation}
\frac{d\beta(t)}{dt} = -\frac{dU(\beta)}{d\beta}+\gamma
cos(\omega_0 t) + \xi_{\beta}(t) \label{beta_eq},
\end{equation}
where $U(\beta)$ is a bistable potential (see Fig.1)

\begin{equation}
U(\beta) = h(\beta-(1+\rho))^4/\eta^4-2h(\beta-(1+\rho))^2/\eta^2,
\label{U(beta)}
\end{equation}
and $h$ is the height of the potential barrier. The periodic term
takes into account for the environment temperature variation. Here
$\gamma=10^{-1}$ and $\omega_0/(2\pi)=10^{-3}$. In
Eq.(\ref{beta_eq}) $\xi_{\beta}(t)$ is a Gaussian white noise with
the usual statistical properties: $\langle
\xi_{\beta}(t)\rangle=0$ and $\langle
\xi_{\beta}(t)\xi_{\beta}(t')\rangle = \sigma\delta(t-t')$. Due to
the shape of $U(\beta)$ it is reasonable to expect a coexistence
regime for $\beta(0) < 1 $, when deterministic case
($\xi_{\beta}(t)=0$) is considered.

\subsection{Stochastic resonance}

First we investigate the effect of the noise on the time behavior
of the species. Since the dynamics of the species strongly depends
on the value of the interaction parameter, we initially analyze
the time evolution of $\beta(t)$ for different levels of the
additive noise $\sigma_\beta$.
\begin{figure}[htbp]
\begin{center}
\includegraphics[width=14cm]{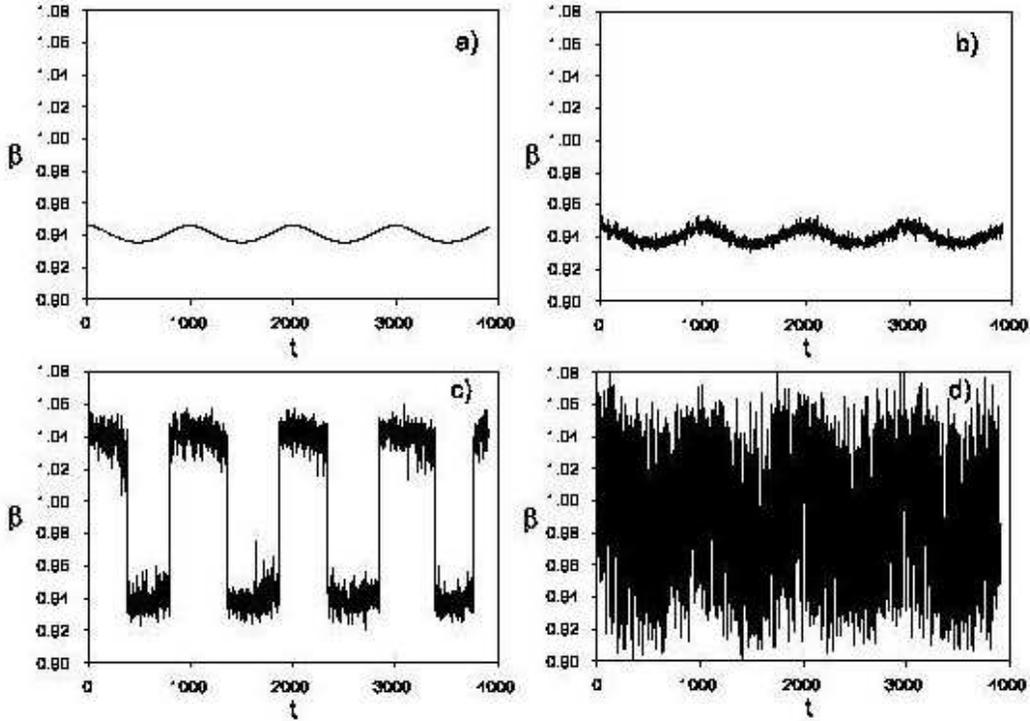}
\end{center}
\caption{ \small \emph{Time evolution of the interaction parameter
for different values of the additive noise $\sigma_\beta$. (a)
$\sigma_\beta=0$; (b) $\sigma_\beta=1.78\cdot 10^{-4}$; (c)
$\sigma_\beta=1.78\cdot 10^{-3}$; (d) $\sigma_\beta=1.78\cdot
10^{-2}$. The values of the parameters are: $\gamma=10^{-1}$,
$\omega_0/(2\pi)=10^{-3}$.}\bigskip} \label{beta_series}
\end{figure}
\begin{figure}[htbp]
\begin{center}
\includegraphics[width=14cm]{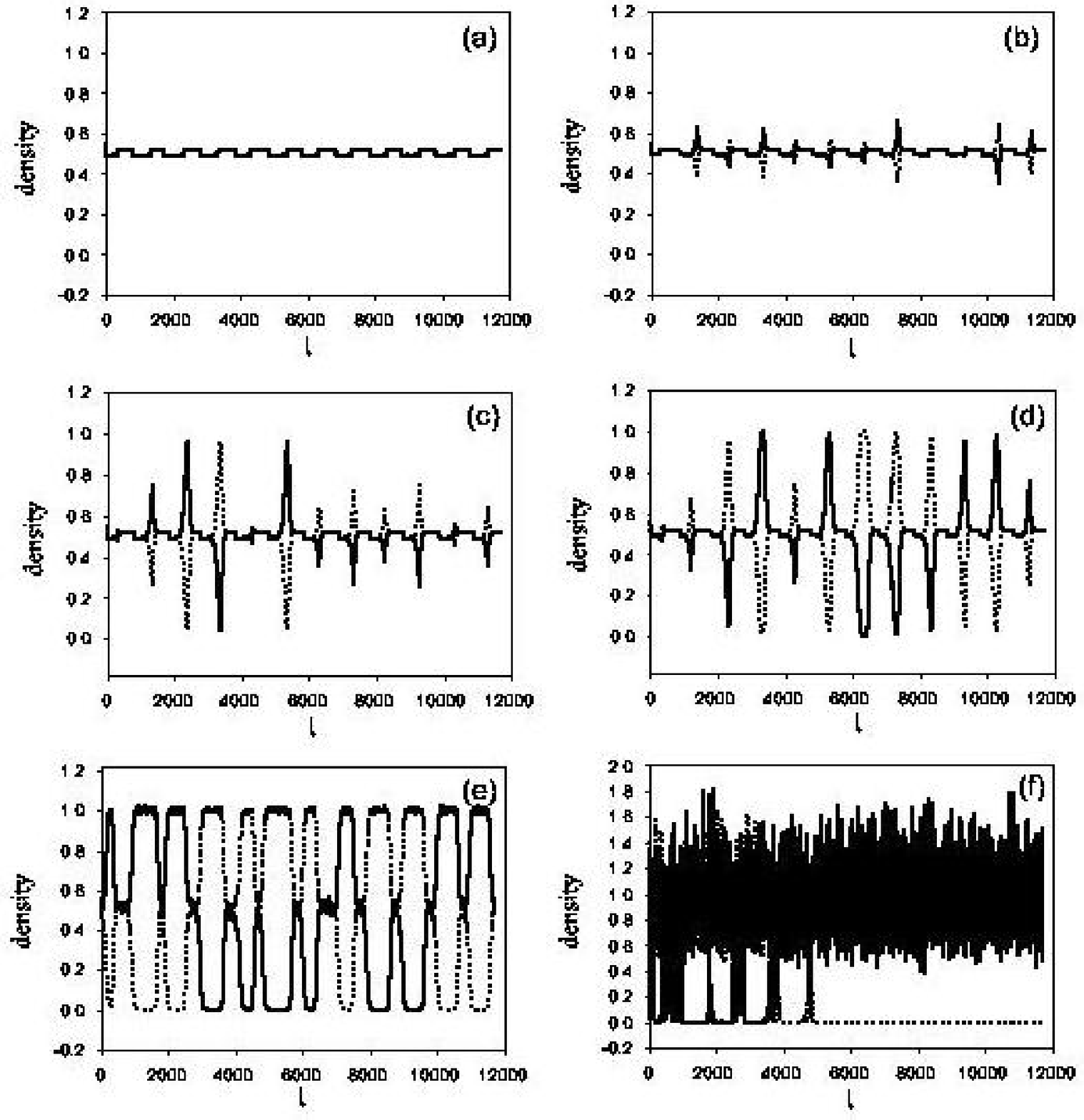}
\end{center}
\caption{ \small \emph{Time evolution of both populations at
different levels of the multiplicative noise: (a) $\sigma=0$; (b)
$\sigma=10^{-11}$; (c) $\sigma=10^{-10}$; (d) $\sigma=10^{-9}$;
(e) $\sigma=10^{-4}$; (f) $\sigma=10^{-1}$. The values of the
parameters are $\mu = 1$, $\alpha=1$, $\gamma = 10^{-1}$,
$\w0/2\pi = 10^{-3}$. The intensity of the additive noise is fixed
at the value $\sigma_\beta=1.78 \cdot 10^{-3}$. The initial values
of the two species are $x(0)=y(0)=1$.}\bigskip}
\label{time_series}
\end{figure}
Specifically for $\sigma_\beta=0$ we obtain a periodical behavior
of $\beta(t)$ in the coexistence region (see
Fig.\ref{beta_series}a). For low noise intensity ($\sigma_\beta
\ll h$) we get the same periodical behavior of deterministic case,
slightly perturbed by the noise (Fig.\ref{beta_series}b). For
higher noise intensity ($\sigma_\beta \simeq h $) the periodical
behavior of the interaction parameter jumps between the two values
$\beta=0.94<1$ and $\beta=1.04>1$, which characterize the
coexistence and the exclusion regimes. We see in
Fig.(\ref{beta_series}c) the typical picture of stochastic
resonance. A further increase of the noise intensity produces a
loss of coherence and the dynamical behavior is strongly
controlled by the noise (Fig.\ref{beta_series}d). There is
therefore a cooperation between periodical driving of the
temperature, due to some geological cause, and the environmental
noise. The noise can synchronize with the periodical driving
giving rise to the stochastic resonance effect, which affect
considerably the dynamics of the ecosystem. The exact value of
$\sigma_\beta=1.78\cdot 10^{-3}$ used in Fig.\ref{beta_series}c is
obtained from the statistical synchronization typical of the SR
phenomenon

\begin{equation}
\tau_k=T_0/2 \label{SR_sync}
\end{equation}
where $\tau_k$ is the Kramers time given by

\begin{equation}
\tau_k = \frac{2 \pi}{\sqrt{\vert U''(0.99) \vert U''(0.94)}}
\exp{[2 h/\sigma_\beta]} . \label{kramers}
\end{equation}
and $T_0$ is the period of the driving force. In
Eq.(\ref{kramers}) $U''(0.99)$ and $U''(0.94)$ are the second
derivative respectively calculated in the unstable and stable
states of the potential. To analyze the dynamics of the two
species we fix the additive noise intensity at the value
$\sigma_\beta=1.78\cdot 10^{-3}$, corresponding to a competition
regime between the two species periodically switched from
coexistence to exclusion. The temporal series of the two species
are obtained for different values of the multiplicative noise
intensity $\sigma=\sigma_x=\sigma_y$. The initial values of the
two species are x(0) = y(0) = 1. After a short transient in the
coexistence regime, which is not visible in Fig.\ref{time_series}
because of the scale used for the x axes, both species reach the
value: $x_{st} = y_{st} = \alpha/(1+\beta) \approx 1/2$. From then
on the densities of both species oscillate quasi regularly around
this value with amplitudes dependent on the multiplicative noise
intensity. In particular for $\sigma=0$ (see
Fig.\ref{time_series}a) and for very low levels of multiplicative
noise ($\sigma=10^{-12}$) the two species coexist and the
populations show correlated oscillations. For higher level of the
multiplicative noise ($\sigma=10^{-4}$) the amplitude of the
oscillations increases, anti-correlated oscillations appear and
random periodical inversions of populations occur (see
Fig.\ref{time_series}c). A further increase of the noise
($\sigma=10^{-2}$) produces a degradation of the signal and a loss
of coherence of the temporal series for the species (see
Fig.\ref{time_series}d). In the absence of multiplicative noise
the Lotka-Volterra equations are symmetric. The choice of the same
values for the initial conditions of the two species densities
maintains this symmetry, so the species oscillate in phase around
the stationary value $\alpha/(1+\beta)$, even if the additive
noise that controls the SR effect of the interaction parameter was
fixed at a value that produces many transitions to the exclusion
regime (see Fig.\ref{time_series}a). In the presence of  very weak
multiplicative noise ($\sigma=10^{-12}$) the species oscillate in
phase almost regularly with some little anticorrelated spikes. By
increasing the amplitude of the multiplicative noise we obtain an
enhancement of the amplitude of both species, as we can see from
Figs. 3b, 3c and 3d. These figures show this transient behaviour
from a regime of quasi correlated oscillations
(Fig.\ref{time_series}b) to a regime of anticorrelated
oscillations (Fig.\ref{time_series}d). The presence of the
multiplicative noise breaks the symmetric dynamical behaviour of
the ecosystem. For $\beta>1$, i. e. in the exclusion regime, this
symmetry breaking determines the "divergent" behaviour of the
trajectories of the two species: one tends to survive, while the
other one tends to extinguish. From the inspection of
Fig.\ref{time_series} we note that in the presence of a bistable
potential modulated by a weak periodic force the response of the
system may be enhanced by the presence of the noise. The
periodicity of the noise-induced oscillations in the time behavior
of the population densities shown in Fig.\ref{time_series}e is the
same of the driving periodic term of Eq.(\ref{beta_eq}). This is
the signature of stochastic resonance (SR). In order to underline
the presence of SR we analyze the squared difference of population
densities $(x-y)^2$. In Fig.\ref{snr} it is shown the SNR of this
quantity as a function of the multiplicative noise intensity
$\sigma$, for $\sigma_\beta=1.78 \cdot 10^{-3}$. We calculate the
SNR of $(x-y)^2$ because of the random periodical inversion of
population densities. We note that dynamics of $(x-y)$ is mainly
affected by the multiplicative noise, as we can see from
Eqs.(\ref{LotVol}). The SNR has been obtained by performing $190$
realizations of Eqs. (\ref{LotVol}). A maximum at $\sigma=10^{-4}$
is present. From the above analysis it is clear the role of the
two noise sources: the additive noise determines the conditions
for the different dynamical regimes of the two species, the
multiplicative noise produces a coherent response of the system by
a mechanism of symmetry breaking of the dynamical evolution of the
ecosystem.
\begin{figure}[htbp]
\begin{center}
\includegraphics[width=11cm]{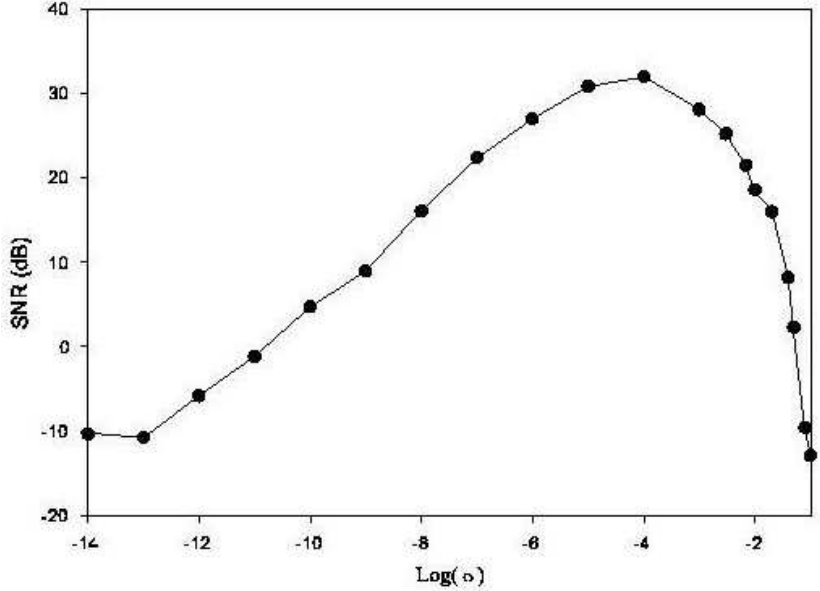}
\end{center}
\caption{ \small \emph{Log-Log plot of SNR as a function of the
multiplicative noise intensity. The SNR corresponds to the squared
difference of population densities $(x - y)^2$. The values of the
parameters are the same of Fig.(\ref{time_series}).}\bigskip}
\label{snr}
\end{figure}
\subsection{Noise delayed extinction}

Now we investigate the average extinction time of one species as a
function of the noise intensity $\sigma_\beta $. To do this we fix
the multiplicative noise at a low level intensity in such a way
that the system is far enough from the SR regime and the dynamics
is not strongly perturbed by the noise.
\begin{figure}[htbp]
\begin{center}
\includegraphics[width=14cm]{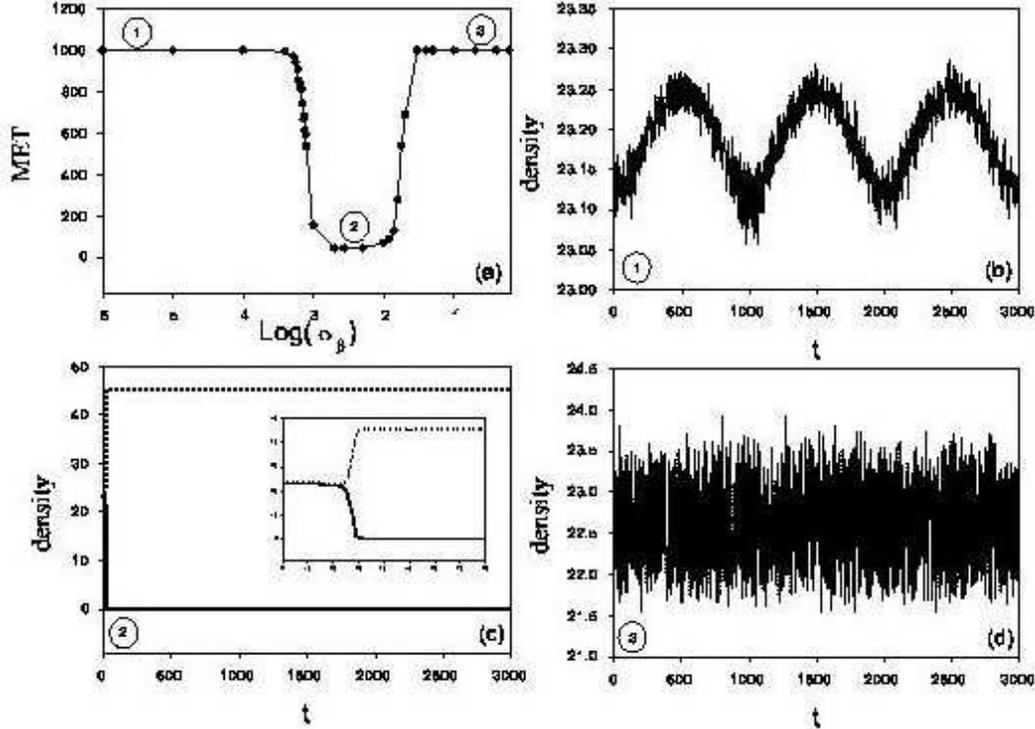}
\end{center}
\caption{\small \emph{(a) Mean extinction time of one species as a
function of the noise intensity $\sigma_\beta$. Time evolution of
both species for different levels of additive noise: (b)
$\sigma_\beta=10^{-4}$, (c) $\sigma_\beta=2 \cdot 10^{-3}$, (d)
$\sigma_\beta=10^{-1}$. The values of the parameters are $\mu =
1$, $\alpha=45$, $\gamma = 10^{-1}$, $\w0/2\pi = 10^{-3}$. The
intensity of the multiplicative noise is fixed at the value
$\sigma=10^{-9}$. The initial values of the two species are
$x(0)=y(0)=1$.}\bigskip} \label{met}
\end{figure}
So we choose $\sigma=10^{-9}$ because we are not interested in the
coherent behavior of the ecological system but we are focused on
the effect of the additive noise on the average extinction time of
the species. We do simulations of Eqs. (1) and (2) by performing
200 realizations. The results are shown in Fig.(\ref{met}). We use
the bistable potential $U(\beta)$ of Fig.\ref{potential} with the
initial condition $\beta(0)=0.94$. In this condition the ecosystem
is in the coexistence regime, that is the deterministic extinction
time of both species is infinite. By introducing noise, exclusion
takes place and a finite mean extinction time (MET) appears. By
varying the intensity of the additive noise in Eq.\ref{beta_eq} we
obtain, of course, a variation of the average extinction time. The
delayed extinction is obtained for noise intensities ranging from
the intermediate regime (2 in Fig.\ref{met}a) to the coexistence
regime obtained with higher values of $\sigma$ (3 in
Fig.\ref{met}a). In real ecosystems we have always a given noise
intensity, which corresponds to a finite mean extinction time. Due
to some environmental cause the noise intensity can considerably
change, as it is observed in experimental data of populations in a
very long time interval~\cite{Car}. Therefore the dynamical
behaviour shown in Fig.\ref{met} should explain such physical
situations, where the variation of the environmental noise
produces a delayed extinction of some population. By increasing
the noise intensity we obtain noise delayed extinction and the
average extinction time grows reaching a saturation value, which
corresponds to a situation where the potential barrier is absent.
We find nonmonotonic behaviour of the MET as a function of the
noise intensity $\sigma_{\beta}$, with a minimum value
$\tau_{min}=40.47$ at $\sigma_{\beta}=2.75 \cdot 10^{-3}$, which
is of the same order of magnitude of the barrier height $h$ (see
Fig.\ref{met}a). The Kramers time corresponding to this noise
intensity is $\tau_k=41.6$, that is approximately equal to
$\tau_{min}$. This result is due to the noise driven dynamics. In
fact for low value of the noise intensities the average time to
overcome the potential barrier is very high, i.e. long Kramers
times. The ecosystem remains in the coexistence regime for a long
time and the extinction time is very large. For noise intensity of
the same order of magnitude of the barrier height the system goes
towards the exclusion regime of one of two species and the average
extinction time is approximately equal to the Kramers time. We get
the minimum value of MET.
\begin{figure}[htbp]
\begin{center}
\includegraphics[width=12cm]{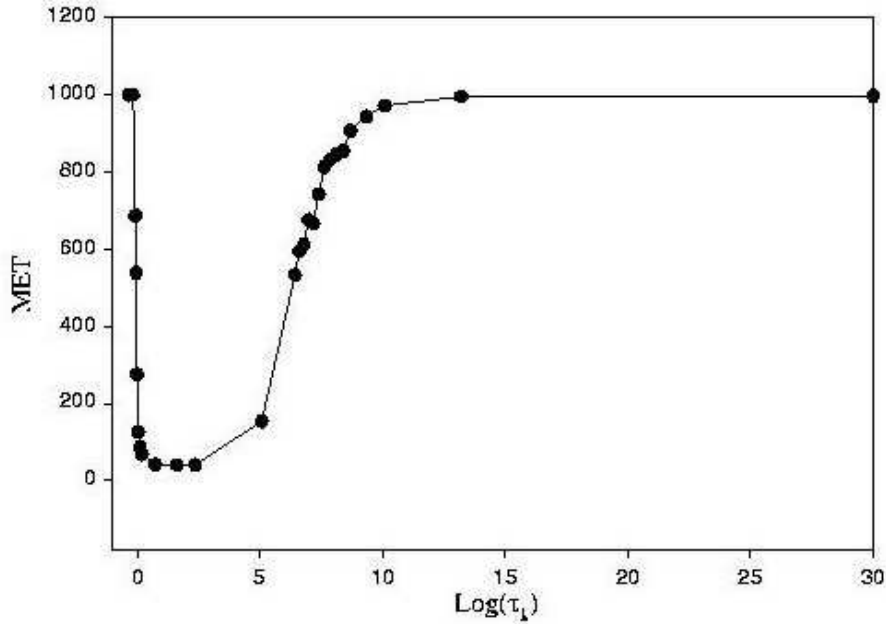}
\end{center}
\caption{ \small \emph{Semilog plot of the mean extinction time of
one species as a function of the Kramers time.}}
\label{met_kramers}
\end{figure}
For higher values of noise intensity the Kramers time becomes very
small and the representative point of the $\beta$ parameter moves
between the two minima in a very short time. In this condition the
system "sees" the average value of the interaction parameter
($\beta=0.99$), which gives a coexistence regime. In
Figs.\ref{met}b,c,d we show the time evolution of the ecosystem
corresponding to the points $1$, $2$ and $3$ of Fig.\ref{met}a. We
have coexistence regime in points $1$ and $3$ and an exclusion
regime in point $2$. We report finally in the following
Fig.\ref{met_kramers} the behavior of the average extinction time
as a function of the Kramers time. This figure confirms our
physical description of the extinction dynamics of the system.

\section{Conclusions}
We report a study on the role of the noise in the dynamics of two
competing species. We consider two noise sources: a multiplicative
noise and an additive noise, which produces a random interaction
parameter between the species. The noise induces a coherent time
behavior of two species giving rise to temporal oscillations and
enhancement of the response of the system through stochastic
resonance phenomenon. Specifically the additive noise controls the
switching between the coexistence and the exclusion dynamical
regimes, the multiplicative noise is responsible for coherent
oscillations of the two species. The SR in the dynamics of
interaction parameter $\beta$ induces SR phenomenon in two
competing species. These time behaviors are absent in the
deterministic dynamics. We note that our model is useful to
describe physical situations in which the amplitude of the
periodical driving force, due to the temperature variations, is
weak and therefore unable to produce considerable variations of
the dynamical regime of the ecosystem. The synergetic cooperation
between the nonlinearity of the system and the random and
periodical environmental driving forces produces therefore a
coherent time behaviour of the ecosystem investigated. The noise
is also responsible for a delayed extinction which gives rise to a
nonmonotonic behavior of the average extinction time as a function
of the additive noise intensity. We note finally that these noise
induced effects should be useful to explain the time evolution of
ecological species, whose dynamics is strongly affected by the
noisy environment~\cite{Spa1,Zim,Gar,Car}.
\section{Acknowledgments}
The authors are grateful to Dr. A. La Barbera for useful
discussions concerning numerical simulations. This work was
supported by INFM, MIUR and INTAS Grant 01-450.

\end{document}